\documentclass[sigconf]{acmart}

\usepackage{booktabs} 
\usepackage{graphicx}
    \setkeys{Gin}{width=5.5cm}
\usepackage{subfigure} 
\usepackage[utf8]{inputenc}

\usepackage{algorithm}
\usepackage{algorithmic}
\usepackage{amssymb}
\usepackage{amsmath}
\usepackage{bm}
\usepackage{mathtools}
\usepackage{paralist}
\usepackage{bbm}
\usepackage{dsfont}
\usepackage{balance}

\graphicspath{ {figures/} }

\copyrightyear{2019}
\acmYear{2019}
\setcopyright{acmcopyright}
\acmConference[SIGIR '19]{Proceedings of the 42nd International ACM SIGIR Conference on Research and Development in Information Retrieval}{July 21--25, 2019}{Paris, France}
\acmBooktitle{Proceedings of the 42nd International ACM SIGIR Conference on Research and Development in Information Retrieval (SIGIR '19), July 21--25, 2019, Paris, France}
\acmPrice{15.00}
\acmDOI{10.1145/3331184.3331202}
\acmISBN{978-1-4503-6172-9/19/07}

\usepackage{etoolbox}
\newtoggle{CRC}
\toggletrue{CRC}

\usepackage{pgf}

\DeclareMathOperator{\relx}{r}

\DeclareMathOperator{\rank}{rank}

\DeclareMathOperator{\argmin}{argmin}
\DeclareMathOperator{\argsort}{argsort}

\DeclareMathOperator{\Prob}{P}

\newcommand{\x}{\bm{x}}                  
\newcommand{\xsample}{\bm{X}}            
\newcommand{\y}{\bm{y}}                  
\newcommand{\ypres}{\bar{\bm{y}}}        
\newcommand{\ypart}{y}                   
\newcommand{\obs}{o}                     
\newcommand{\Loss}{\Delta}               
\newcommand{\Risk}{R}                    
\newcommand{\Losshat}{\hat{\Loss}}       
\newcommand{\Riskhat}{\hat{\Risk}}       
\newcommand{\weight}[1]{\lambda\left(#1\right)}
\newcommand{\dcgneg}[1]{\frac{-1}{\log(1+#1)}}

\newcommand{\E}{{\mathbb{E}}}            
\newcommand{\systems}{{\mathcal{S}}}     
\newcommand{\system}{{S}}                
\newcommand{\n}{{n}}                     
\newcommand{\N}{{N}}                     
\newcommand{\w}{w}                       
\newcommand{\what}{\hat{w}}              

\begin{document}
\title{A General Framework for Counterfactual Learning-to-Rank}

\author{Aman Agarwal}
\affiliation{%
  \institution{Cornell University}
  \city{Ithaca} 
  \state{NY} 
}
\email{aman@cs.cornell.edu}

\author{Kenta Takatsu}
\affiliation{%
  \institution{Cornell University}
  \city{Ithaca} 
  \state{NY} 
}
\email{kt426@cornell.edu}

\author{Ivan Zaitsev}
\affiliation{%
  \institution{Cornell University}
  \city{Ithaca} 
  \state{NY} 
}
\email{iz44@cornell.edu}

\author{Thorsten Joachims}
\affiliation{%
  \institution{Cornell University}
  \city{Ithaca} 
  \state{NY} 
}
\email{tj@cs.cornell.edu}


\begin{abstract}
Implicit feedback (e.g., click, dwell time) is an attractive source of training data for Learning-to-Rank, but its naive use leads to learning results that are distorted by presentation bias. For the special case of optimizing average rank for linear ranking functions, however, the recently developed SVM-PropRank method has shown that counterfactual inference techniques can be used to provably overcome the distorting effect of presentation bias.
Going beyond this special case, this paper provides a general and theoretically rigorous framework for counterfactual learning-to-rank that enables unbiased training for a broad class of additive ranking metrics (e.g., Discounted Cumulative Gain (DCG)) as well as a broad class of models (e.g., deep networks). Specifically, we derive a relaxation for propensity-weighted rank-based metrics which is subdifferentiable and thus suitable for gradient-based optimization. We demonstrate the effectiveness of this general approach by instantiating two new learning methods. One is a new type of unbiased SVM that optimizes DCG -- called SVM PropDCG --, and we show how the resulting optimization problem can be solved via the Convex Concave Procedure (CCP). The other is Deep PropDCG, where the ranking function can be an arbitrary deep network. In addition to the theoretical support, we empirically find that SVM PropDCG significantly outperforms existing linear rankers in terms of DCG. Moreover, the ability to train non-linear ranking functions via Deep PropDCG further improves performance. 

\end{abstract}

%
%


\keywords{Learning to rank, presentation bias, counterfactual inference}

\maketitle

\section{Introduction}
\label{sec:intro}
Implicit feedback from user behavior is an attractive source of data in many information retrieval (IR) systems, especially ranking applications where collecting relevance annotations from experts can be economically infeasible or even impossible (e.g., personal collection search, intranet search, scholarly search). While implicit feedback is often abundant, cheap, timely, user-centric, and routinely logged, it suffers from inherent biases. For example, the position of a result in a search ranking strongly affects how likely it is to be seen by a user and thus clicked. So, naively using click data as a relevance signal leads to sub-optimal performance. 

A counterfactual inference approach for learning-to-rank (LTR) from logged implicit feedback was recently developed to deal with such biases \cite{Joachims/etal/17a}. This method provides a rigorous approach to unbiased learning despite biased data and overcomes the limitations of alternative bias-mitigation strategies. In particular, it does not require the same query to be seen multiple times as necessary for most generative click models, and it does not introduce alternate biases like treating clicks as preferences between clicked and skipped documents. 

The key technique in counterfactual learning is to incorporate the propensity of obtaining a particular training example into an Empirical Risk Minimization (ERM) objective that is provably unbiased \cite{Swaminathan/Joachims/15c}. While it was shown that this is possible for learning to rank, existing theoretical support is limited to linear ranking functions and optimizing average rank of the relevant documents as objective \cite{Joachims/etal/17a}. In this paper, we generalize the counterfactual LTR framework to a broad class of additive IR metrics as well as non-linear deep models. Specifically, we show that any IR metric that is the sum of individual document relevances weighted by some function of document rank can be directly optimized via Propensity-Weighted ERM. Moreover, if an IR metric meets the mild requirement that the rank weighting function is monotone, we show that a hinge-loss upper-bounding technique enables learning of a broad class of differentiable models (e.g. deep networks). 

To demonstrate the effectiveness of the general framework, we fully develop two learning-to-rank methods that optimize the Discounted Cumulative Gain (DCG) metric. The first is SVM PropDCG, which generalizes a Ranking SVM to directly optimize a bound on DCG from biased click data. The resulting optimization problem is no longer convex, and we show how to find a local optimum using the Convex Concave Procedure (CCP). In CCP, several iterations of convex sub-problems are solved. In the case of SVM PropDCG, these convex sub-problems have the convenient property of being a Quadratic Program analogous to a generalized Ranking SVM. This allows the CCP to work by invoking an existing and fast SVM solver in each iteration until convergence. The second method we develop is Deep PropDCG, which further generalizes the approach to deep networks as non-linear ranking functions. Deep PropDCG also optimizes a bound on the DCG, and we show how the resulting optimization problem can be solved via stochastic gradient descent for any network architecture that shares neural network weights across candidate documents for the same query.  

In addition to the theoretical derivation and the justification, we also empirically evaluate the effectiveness of both SVM PropDCG and Deep PropDCG, especially in comparison to the existing SVM PropRank method \cite{Joachims/etal/17a}. We find that SVM PropDCG performs significantly better than SVM PropRank in terms of DCG, and that it is robust to varying degrees of bias, noise and propensity-model misspecification. In our experiments, CCP convergence was typically achieved quickly within three to five iterations. For Deep PropDCG, the results show that DCG performance is further improved compared to SVM PropDCG when using a neural network, thus demonstrating that the counterfactual learning approach can effectively train non-linear ranking functions. SVM PropDCG and Deep PropDCG are also seen to outperform LambdaRank in terms of DCG. 



\section{Related Work}
\label{sec:related}
Generative click models are a popular approach for explaining the bias in user behavior and for extracting relevance labels for learning. For example, in the cascade model \cite{craswell2008position} users are assumed to sequentially go down a ranking and click on a document, thus revealing preferences between clicked and skipped documents.
Learning from these relative preferences lowers the impact of some biases \cite{Joachims/02c}. 
Other 
click models (\cite{craswell2008position,Chapelle/Zhang/09,borisov2016neural}, also see \cite{Chuklin/etal/15a}) 
train to maximize log-likelihood of observed clicks, where relevance is modeled as a latent variable that is inferred over multiple instances of the same query. In contrast, the counterfactual framework \cite{Joachims/etal/17a} does not require latent-variable inference and repeat queries, but allows directly incorporating click feedback into the learning-to-rank algorithm in a principled and unbiased way, thus allowing the direct optimization of ranking performance over the natural query distribution. 

The counterfactual approach uses inverse propensity score (IPS) weighting, originally employed in survey sampling \cite{Horvitz1952} and causal inference from observational studies \cite{Rosenbaum1983}, but more recently also in whole page optimization \cite{Wang2016}, IR evaluation with manual judgments \cite{Schnabel/etal/16c}, and recommender evaluation \cite{li2011unbiased,Schnabel/etal/16b}. This approach is similar in spirit to  \cite{Wang/etal/16}, where propensity-weighting is used to correct for selection bias when discarding queries without clicks during learning-to-rank. 

Recently, inspired by the IPS correction approach for unbiased LTR, some algorithms (\citet{ai2018unbiased, hu2018novel}) have been proposed  that jointly estimate the propensities and learn the ranking function. However, this requires an accurate relevance model to succeed, which is at least as hard as the LTR from biased feedback problem in question. Moreover, the two-step approach of propensity estimation followed by training an unbiased ranker allows direct optimization of any chosen target ranking metric independent of the propensity estimation step.

While our focus is on directly optimizing ranking performance in the implicit feedback partial-information setting, several approaches have been proposed for the same task in the full-information supervised setting, i.e. when the relevances of all the documents in the training set are known. A common strategy is to use some smoothed version of the ranking metric for optimization, as seen in SoftRank \cite{softrank} and others \cite{Chapelle,Wu,Yue/etal/07a,Joachims/etal/09a}. In particular, SoftRank optimizes the expected performance metric over the distribution of rankings induced by smoothed scores, which come from a normal distribution centered at the query-document  mean scores predicted by a neural net. This procedure is computationally expensive with an $O(n^3)$ dependence on the number of documents for a query. In contrast, our approach employs an upper bound on the performance metric, whose structure makes it amenable to the Convex Concave Procedure for efficient optimization, as well as adaptable to non-linear ranking functions via deep networks.   

Finally, several works exist \cite{softrank, burgesltr, Burgesltr2, sortnet} that have proposed neural network architectures for learning-to-rank. We do not focus on a specific network architecture in this paper, but instead propose a new training criterion for learning-to-rank from implicit feedback that in principle allows unbiased network training for a large class of architectures.  
\section{Unbiased Estimation of Rank-Based IR Metrics}
\label{sec:ltr}

\begin{table}[tb]
  \centering
    \begin{tabular}{lll}
    \toprule
    Metric &  $\weight{\mathrm{rank}}$ \\
    \midrule
    $Avg Rank$ & $\mathrm{rank}$ \\
    $DCG$ & $ -1\!/\!\log\!( 1 + \mathrm{rank})$ \\ 
    $Prec@k$ & $-\mathbf{1}_{\mathrm{rank} \leq k}/k$ \\
    \emph{RBP-p} \cite{Moffat2008} &  $-(1-p)/p^\mathrm{rank}$ \\
    \bottomrule
    \end{tabular}%
  \caption{Some popular linearly decomposable IR metrics that can be directly optimized by Propensity-Weighted ERM. $\weight{r}$ is the rank weighting function.}
  \vspace{-0.8em}
  \label{tbl:metrics}%
\end{table}%

We begin by developing a counterfactual learning framework that covers the full class of linearly decomposable metrics as defined below (e.g. DCG). This extends \cite{Joachims/etal/17a} which was limited to the Average Rank metric. 
Suppose we are given a sample $\xsample$ of i.i.d. query instances $\x_i \sim \Prob(\x)$, $i \in [N]$. A query instance can include personalized and contextual information about the user in addition to the query string. For each query instance $\x_i$, let $\relx_{i}(\ypart)$ denote the user-specific relevance of result $\ypart$ for instance $\x_i$. For simplicity, assume that relevances are binary, $\relx_{i}(\ypart) \in \{0,1\}$. In the following, we consider the class of additive ranking performance metrics, which includes any metric that can be expressed as
\begin{eqnarray}
    \Loss(\y|\x_i,\relx_i) & = & \sum_{\ypart \in \y} \weight{\rank(\ypart|\y)} \cdot \relx_i(\ypart). \label{eq:metricfullinfo}
\end{eqnarray}
$\y$ denotes a ranking of results, and $\weight{}$ can be any weighting function that depends on the rank $\rank(\ypart|\y)$ of document $\ypart$ in ranking $\y$. A broad range of commonly used ranking metrics falls into this class, and Table~\ref{tbl:metrics} lists some of them. For instance, setting $\weight{\mathrm{rank}} = \mathrm{rank}$ gives the sum of relevant ranks metric (also called average rank when normalized) considered in \cite{Joachims/etal/17a}, and $\weight{\mathrm{rank}} = \dcgneg{\mathrm{rank}}$ gives the DCG metric. Note that we consider negative values wherever necessary to make the notation consistent with risk minimization. 

A ranking system $\system$ maps a query instance $\x_i$ to a ranking $\y$. Aggregating the losses of individual rankings over the query distribution, we can define the overall \emph{risk} (e.g., the expected DCG) of a system as 
\begin{eqnarray}
    \Risk(\system) & = & \int \Loss(\system(\x)|\x,\relx) \: d \Prob(\x,\relx). \label{eq:utilpartialinfo}
\end{eqnarray}

A key problem when working with implicit feedback data is that we cannot assume that all relevances $\relx_i$ are observed. In particular, while a click (or a sufficiently long dwell time) provides a noisy indicator of positive relevance in the presented ranking $\ypres_i$, a missing click does not necessarily indicate lack of relevance as the user may not have observed that result. From a machine learning perspective, this implies that we are in a partial-information setting, which we will deal with by explicitly modeling missingness in addition to relevance. Let $\obs_i \sim \Prob(\obs|\x_i, \ypres_i,\relx_i)$ denote the 0/1 vector indicating which relevance values are revealed. While $\obs_i$ is not necessarily fully observed either, we can now model its distribution, which we will find below is sufficient for unbiased learning despite the missing data. In particular, the \emph{propensity} of observing $\relx_i(\ypart)$ for query instance $x_i$ given presented ranking $\ypres$ is then defined as $Q(\obs_i(\ypart)=1|\x_i,\ypres_i, \relx_i)$.

Using this counterfactual setup, an unbiased estimate of $\Loss(\y|\x_i,\relx_i)$ for any ranking $\y$ can be obtained via IPS weighting 
\begin{eqnarray}
    \Losshat_{IPS}(\y|\x_i,\ypres_i, \obs_i) & \!\!\!=\!\!\! & \sum_{\substack{ \ypart : \obs_i(\ypart) = 1 \\
    \bigwedge \relx_i(\ypart) = 1}}  \frac{\weight{\rank(\ypart|\y)} }{Q(\obs_{i}(\ypart)\!=\!1|\x_i,\ypres_i,\relx_i)}. \label{eq:lossips}
\end{eqnarray}
This is an unbiased estimator since,
\begin{eqnarray}
&&\!\!\!\!\!\!\!\!\!\!\!\!\!\!\!\!\E_{\obs_i}[\Losshat_{IPS}(\y|\x_i,\ypres_i, \obs_i)]  \\
    & = & \E_{\obs_i}\!\!\left[ \sum_{\ypart : \obs_i(\ypart) = 1}  \frac{\weight{\rank(\ypart|\y)} \!\cdot\! \relx_i(\ypart)}{Q(\obs_{i}(\ypart)\!=\!1|\x_i,\ypres_i,\relx_i)} \right] \nonumber \\
    & = & \sum_{\ypart \in \y} \E_{\obs_i}\!\!\left[  \frac{\obs_i(\ypart) \!\cdot\! \weight{\rank(\ypart|\y)} \!\cdot\! \relx_i(\ypart)}{Q(\obs_{i}(\ypart)\!=\!1|\x_i,\ypres_i,\relx_i)} \right] \nonumber \\
    & = & \sum_{\ypart \in \y} \frac{Q(\obs_i(\ypart)=1|\x_i,\ypres_i,\relx_i) \cdot \weight{\rank(\ypart|\y)} \cdot \relx_i(\ypart) }{Q(\obs_i(\ypart)=1|\x_i,\ypres_i,\relx_i)}  \nonumber \\
    & = & \sum_{\ypart \in \y} \weight{\rank(\ypart|\y)} \relx_i(\ypart)  \nonumber \\
    & = & \Loss(\y|\x_i,\relx_i),\nonumber
\end{eqnarray}
 assuming $Q(\obs_i(\ypart)=1|\x_i,\ypres_i,\relx_i) > 0$ for all $\ypart$ that are relevant $\relx_i(\ypart) = 1$. The above proof is a generalized version of the one in \cite{Joachims/etal/17a} for the Average Rank metric. Note that the estimator in Equation~\eqref{eq:lossips} sums only over the results where the feedback is observed (i.e., $\obs_i(\ypart)=1$) and positive (i.e., $\relx_i(\ypart) = 1$), which means that we do not have to disambiguate whether lack of positive feedback (e.g., the lack of a click) is due to a lack of relevance or due to missing the observation (e.g., result not relevant vs. not viewed). 

Using this unbiased estimate of the loss function, we get an unbiased estimate of the risk of a ranking system $\system$
\begin{eqnarray}
    \!\!\!\!\Riskhat_{IPS}(\system) \!\!\!\!& = \!\!\!\!& \frac{1}{\N}\!\sum_{i=1}^{\N} \sum_{\substack{ \ypart : \obs_i(\ypart) = 1 \\
    \bigwedge \relx_i(\ypart) = 1}}  \frac{\weight{\rank(\ypart|\system(\x_i))} }{Q(\obs_{i}(\ypart)\!=\!1|\x_i,\ypres_i,\relx_i)} . \:\:\:\: \label{eq:riskips}
\end{eqnarray}

Note that the propensities $Q(\obs_{i}(\ypart)\!=\!1|\x_i,\ypres_i,\relx_i)$ are generally unknown, and must be estimated based on some model of user behavior. Practical approaches to estimating the propensities are given in \cite{Joachims/etal/17a, Wang/etal/18, Agarwal/etal/19, Fang/etal/19a}. 

\section{Unbiased Empirical Risk Minimization for LTR}
The propensity-weighted empirical risk from Equation~\eqref{eq:riskips} can be used to perform Empirical Risk Minimization (ERM)
\begin{eqnarray}
    \hat{\system} & = & \argmin_{\system \in \systems} \left\{\Riskhat_{IPS}(\system) \right\}. \nonumber
\end{eqnarray}
Under the standard uniform convergence conditions \cite{Vapnik1998}, the unbiasedness of the risk estimate implies consistency in the sense that given enough training data, the learning algorithm is guaranteed to find the best system in $\systems$. We have thus obtained a theoretically justified training objective for learning-to-rank with additive metrics like DCG. However, it remains to be shown that this training objective can be implemented in efficient and practical learning methods. This section shows that this is indeed possible for a generalization of Ranking SVMs and for deep networks as ranking functions.

Consider a dataset of $\n$ examples of the following form. For each query-result pair $(\x_i,\ypart_i)$ that is clicked, let $q_i=Q(\obs_{i}(\ypart)=1|\x_i,\ypres_i,\relx_i)$ be the propensity of the click according to a click propensity model such as the Position-Based Model \cite{Joachims/etal/17a, Wang/etal/18}. We also record the candidate set $Y_i$ of all results for query $\x_i$. Note that each click generates a separate training example, even if multiple clicks occur for the same query.

Given this propensity-scored click data, we would like to learn a scoring function $f(\x,\ypart)$. Such a scoring function $f$ naturally specifies a ranking system $\system$ by sorting candidate results $Y$ for a given query $\x$ by their scores. 
\begin{eqnarray}
    \system_f(\x) \equiv \argsort_Y{\left\{f(\x,\ypart)\right\}}
\end{eqnarray}
Since $\rank(\ypart|\system_f(\x))$ of a result is a discontinuous step function of the score, tractable learning algorithms typically optimize a substitute loss that is (sub-)differentiable \cite{Joachims/02c,Yue/etal/07a,softrank}. Following this route, we now derive a tractable substitute for the empirical risk of \eqref{eq:riskips} in terms of the scoring function. This is achieved by the following hinge-loss upper bound \cite{Joachims/etal/17a} on the rank
\begin{align*}
    \rank(\ypart_i|\y) - 1 & =  \sum_{\substack{\ypart \in Y_i \\ \ypart \not= \ypart_i}} \mathds{1}_{f(\x_i,\ypart) - f(\x_i,\ypart_i) > 0}\\
    & \leq  \sum_{\substack{\ypart \in Y_i \\ \ypart \not= \ypart_i}} \max(1 - (f(\x_i,\ypart_i) - f(\x_i,\ypart)), 0).
\end{align*}
Using this upper bound, we can also get a bound for any IR metric that can be expressed through a monotonically increasing weighting function $\weight{r}$ of the rank. Note that this monotonicity condition is satisfied by all the metrics in Table~\ref{tbl:metrics}. By rearranging terms and applying the weighting function $\weight{r}$, we have 
\begin{align*}
    \weight{\rank(\ypart_i|\y)} \leq \weight{1+\!\!\sum_{\substack{\ypart \in Y_i \\ \ypart \not= \ypart_i}} \max(1 - (f(\x_i,\ypart_i) - f(\x_i,\ypart)), 0)}.
\end{align*}
This provides the following continuous and subdifferentiable upper bound $\Riskhat_{IPS}^{hinge}(f)$ on the propensity-weighted risk estimate of $\eqref{eq:riskips}$.
\begin{eqnarray}
    \!\!\!\!\!\!\!\!\Riskhat_{IPS}(\system_f) \!\!\!\!\!& \le & \!\!\!\!\!\Riskhat_{IPS}^{hinge}(f) \nonumber \\
    & = &\!\!\!\!\!\! \frac{1}{n} \!\sum_{i=1}^{n} \!\frac{1}{q_i} \weight {\!1\!+\!\!\!\sum_{\substack{\ypart \in Y_i \\ \ypart \not= \ypart_i}} \!\!\max(1 - ( f(\x_i,\ypart_i) \!-\! f(\x_i,\ypart)), 0)}  \:\: \label{eq:riskhinge}
\end{eqnarray}
Focusing on the DCG metric, we show in the following how this upper bound can be optimized for linear as well as non-linear neural network scoring functions. For the general class of additive IR metrics, the optimization depends on the properties of the weighting function $\weight{r}$, and we highlight them wherever appropriate. 

\subsection{SVM PropDCG}

The following derives an SVM-style method, called SVM PropDCG, for learning a linear scoring function $f(\x,\ypart)=\w \cdot \phi(\x,\ypart)$, where $\w$ is a weight vector and $\phi(\x,\ypart)$ is a feature vector describing the match between query $\x$ and result $\ypart$. For such linear ranking functions -- which are widely used in Ranking SVMs \cite{Joachims/02c} and many other learning-to-rank methods \cite{Liu/09} --, the propensity-weighted ERM bound from Equation~\eqref{eq:riskhinge} can be expressed as the following SVM-type optimization problem.
\begin{eqnarray*}
    \what & \!\!=\!\! & \argmin_{w,\xi} \: \frac{1}{2} \w \cdot \w + \frac{C}{n} \sum_{i=1}^n \frac{1}{q_i} \weight{\sum_{\ypart \in Y_i} \xi_{i\ypart} + 1} \\
     s.t. &   & \forall \ypart \in Y_1 \!\setminus\! \{\ypart_1\}: \w \cdot [\phi(\x_1,\ypart_1) - \phi(\x_1,\ypart)] \ge 1\!-\!\xi_{1\ypart} \\
          &   & \hspace{1cm} \vdots \\
          &   & \forall \ypart \in Y_n \!\setminus\! \{\ypart_n\}: \w \cdot [\phi(\x_n,\ypart_n) - \phi(\x_n,\ypart)] \ge 1\!-\!\xi_{n\ypart} \\
          &   & \forall i \forall \ypart: \xi_{i\ypart} \geq 0
\end{eqnarray*}
$C$ is a regularization parameter. The training objective optimizes the $\mathcal{L}_2$-regularized hinge-loss upper bound on the empirical risk estimate \eqref{eq:riskhinge}. This upper bound holds since for any feasible ($\w,\xi$) and any monotonically increasing weighting function $\weight{r}$ 
\begin{align*}
    & \weight{1+\!\!\sum_{\substack{\ypart \in Y_i \\ \ypart \not= \ypart_i}} \max(1 - (f(\x_i,\ypart_i) - f(\x_i,\ypart)), 0)} \\ 
    & = \weight{\!1\!+\!\!\!\sum_{\substack{\ypart \in Y_i \\ \ypart \not= \ypart_i}} \max(1 - \w \cdot [\phi(\x_i,\ypart_i) - \phi(\x_i,\ypart)], 0)} 
     \leq \weight{\!1\!+\!\!\!\sum_{\ypart \in Y_i} \xi_{i\ypart}}.
\end{align*}
As shown in \cite{Joachims/etal/17a}, for the special case of using the sum of relevant ranks as the metric to optimize, i.e. $\weight{r}=r$, this SVM optimization problem is a convex Quadratic Program which can be solved efficiently using standard SVM solvers, like SVM-rank \cite{Joachims/etal/09a}, via a one-slack formulation.

Moving to the case of DCG as the training metric via the weighting function $\weight{r} = \dcgneg{r}$, we get the following optimization problem for SVM PropDCG
\begin{eqnarray*}
    \what & \!\!=\!\! & \argmin_{w,\xi} \: \frac{1}{2} \w \cdot \w - \frac{C}{n} \sum_{i=1}^n \frac{1}{q_i} \frac{1}{\log(\sum_{\ypart \in Y_i} \xi_{i\ypart} + 2)} \\
     s.t. &   & \forall j \forall \ypart \in Y_i \!\setminus\! \{\ypart_i\}: \w \cdot [\phi(\x_i,\ypart_i) - \phi(\x_i,\ypart)] \ge 1\!-\!\xi_{i\ypart} \\
          &   & \forall j \forall \ypart: \xi_{i\ypart} \geq 0.
\end{eqnarray*}
This optimization problem is no longer a convex Quadratic Program. However, all constraints are still linear inequalities in the variables $\w$ and $\xi$, and the objective can be expressed as the difference of two convex functions $h$ an $g$. Let $h(\w) = \frac12\lVert\w\rVert^2$ and $g(\xi) = \frac C\n \sum_{j=1}^{\n} \frac{1}{q_i} \frac{1}{\log(\sum_{y \in \mathbf{Y}_i} \xi_{iy} + 2)}$. Then the function $h$ is the $\mathcal{L}_2$ norm of the vector $\w$ and is thus a convex function. As for the function $g$, the function $k:x \mapsto \frac1{\log x}$ is convex as it is the composition of a a convex decreasing function ($x\mapsto\frac1x$) with a concave function ($x\mapsto\log x$). So, since the sum of affine transformations of a convex function is convex, $g$ is convex.

Such an optimization problem is called a convex-concave problem\footnote{More generally, the inequality constraints can also be convex-concave and not just convex} and a local optimum can be obtained efficiently via the Convex-Concave Procedure (CCP) \cite{cvx-ccv}. At a high level, the procedure works by repeatedly approximating the second convex function with its first order Taylor expansion which makes the optimization problem convex in each iteration. The Taylor expansion is first done at some chosen initial point in the feasible region, and then the solution of the convex problem in a particular iteration is used as the Taylor approximation point for the next iteration. It can be shown that this procedure converges to a local optimum \cite{cvx-ccv}.

Concretely, let $w^k$, $\xi^k$ be the solution in the $k^{th}$ iteration. Then, we have the Taylor approximation 
\begin{eqnarray*}
&\hat{g}(\xi;\xi^k)& = g(\xi^k) + \nabla g(\xi^k)^T(\xi - \xi^k)\\
&=& g(\xi^k) - \frac Cn\sum\limits_{j=1}^n\frac1{q_i}\frac{\sum\limits_{y\in {\bf Y}_i}\xi_{iy}-\xi^k_{iy}}{\left(\sum\limits_{y\in{\bf Y}_i}\xi^k_{iy}+2\right)\log^2\left(\sum\limits_{y\in{\bf Y}_i}\xi^k_{iy}+2\right)}
\end{eqnarray*}

Letting $q_i'=q_i\left(\sum\limits_{y\in{\bf Y}_i}\xi^k_{iy}+2\right)\log^2\left(\sum\limits_{y\in{\bf Y}_i}\xi^k_{iy}+2\right)$, and dropping the additive constant terms from $\hat{g}$, we get the following convex program that needs to be solved in each CCP iteration.
\begin{eqnarray*}
    &  &\argmin_{w,\xi} \: \frac{1}{2} \w \cdot \w + \frac{C}{n} \sum_{i=1}^n \frac{1}{q_i'} \sum_{\ypart \in Y_i} \xi_{i\ypart}\\
     s.t. &   & \forall i \forall \ypart \in Y_i \!\setminus\! \{\ypart_i\}: \w \cdot [\phi(\x_i,\ypart_i) - \phi(\x_i,\ypart)] \ge 1\!-\!\xi_{i\ypart} \\
          &   & \forall i \forall \ypart: \xi_{i\ypart} \geq 0
\end{eqnarray*}
Observe that this problem is of the same form as SVM PropRank, the Propensity Ranking SVM for the average rank metric, i.e. $\weight{r}=r$ (with the caveat that $q_i'$ are not propensities). This nifty feature allows us to solve the convex problem in each iteration of the CCP using the fast solver for SVM PropRank provided in \cite{Joachims/etal/17a}. In our experiments, CCP convergence was achieved within a few iterations -- as detailed in the empirical section. For other IR metrics, the complexity and feasibility of the above Ranking SVM optimization procedure will depend on the form of the target IR metric. In particular, if the rank weighting function $\weight{r}$ is convex, it may be solved directly as a convex program. If $\weight{r}$ is concave, then the CCP may be employed as shown for the DCG metric above. 

An attractive theoretical property of SVM-style methods is the ability to switch from linear to non-linear functions via the Kernel trick. In principle, kernelization can be applied to SVM PropDCG as is evident from the representer theorem \cite{Schoelkopf/Smola/02}. Specifically, by taking the Lagrange dual, the problem can be kernelized analogous to \cite{Joachims/02c}. While it can be shown that the dual is convex and strong duality holds, it is not clear that the optimization problem has a convenient and compact form that can be efficiently solved in practice. Even for the special case of the average rank metric, $\weight{r}=r$, the associated kernel matrix $K_{iy,jy'}$ has a size equal to the total number of candidates $\sum_{i=1}^n |Y_i|$ squared, making the kernelization approach computationally infeasible or challenging at best. We therefore explore a different route for extending our approach to non-linear scoring functions in the following.

\subsection{Deep PropDCG}
Since moving to non-linear ranking functions through SVM kernelization is challenging, we instead explore deep networks as a class of non-linear scoring functions. Specifically, we
replace the linear scoring function $f(\x,\ypart)=\w \cdot \phi(\x,\ypart)$ with a neural network
\begin{eqnarray}
    f(\x,\ypart) = NN_\w[\phi(\x,\ypart)]
\end{eqnarray} 
This network is generally non-linear in both the weights $\w$ and the features $\phi(\x,\ypart)$. However, this does not affect the validity of the hinge-loss upper bound from Equation~\eqref{eq:riskhinge}, which now takes the form
\begin{align*}
\frac{1}{n}\!\sum_{j=1}^{n} \!\frac{1}{q_i} \weight {1\!+\!\!\!\sum_{\substack{\ypart \in Y_i \\ \ypart \not= \ypart_i}} \max(1 - ( NN_\w[\phi(\x_i,\ypart_i)] - NN_\w[\phi(\x_i,\ypart)]), 0)}
\end{align*}
During training, we need to minimize this function with respect to the network parameters $\w$. Unlike in the case of SVM PropDCG, this function can no longer be expressed as the difference of a convex and a concave function, since $NN_\w[\phi(\x_i,\ypart_i)]$ is neither convex nor concave in general. Nevertheless, the empirical success of optimizing non-convex $NN_\w[\phi(\x_i,\ypart_i)]$ via gradient descent to a local optimum is well documented, and we will use this approach in the following. This is possible since the training objective is subdifferentiable as long as the weighting function $\weight{r}$ is differentiable. However, the non-linearity of $\weight{r}$ adds a challenge in applying {\em stochastic} gradient descent methods to our training objective, since the objective no longer decomposes into a sum over all $(\x_i,\ypart)$ as in standard network training. We discuss in the following how to handle this situation to arrive at an efficient stochastic-gradient procedure.

For concreteness, we again focus on the case of optimizing DCG via $\weight{r} = \dcgneg{r}$. In particular, plugging in the weighting function for DCG, we get the Deep PropDCG minimization objective
\begin{align*}
\frac{1}{n}\!\sum_{j=1}^{n} \!\frac{-1}{q_i} \log^{-1}\!\left( \!2\!+\!\!\!\!\sum_{\substack{\ypart \in Y_i \\ \ypart \not= \ypart_i}} \!\!\max(1 \!-\! ( NN_\w[\phi(\x_i,\ypart_i)] \!-\! NN_\w[\phi(\x_i,\ypart)]), 0)\!\right)
\end{align*}
to which a regularization term can be added (our implementation uses weight decay). 

Since the weighting function ties together the hinge losses from pairs of documents in a non-linear way, stochastic gradient descent (SGD) is not directly feasible at the level of individual documents. In the case of DCG, since the rank weighting function is concave, one possible workaround is a Majorization-Minimization scheme \cite{Swaminathan/Joachims/15c} (akin to CCP): upper bound the loss function with a linear Taylor approximation at the current neural net weights, perform SGD at the level of document pairs $(\ypart_i,\ypart)$ to update the weights, and repeat until convergence. 

While this Majorization-Minimization scheme in analogy to the SVM approach is possible also for deep networks, we chose a different approach for the reasons given below. In particular, given the success of stochastic-gradient training of deep networks in other settings, we directly perform stochastic-gradient updates at the level of query instances, not individual $(\x_i,\ypart)$. At the level of query instances, the objective does decompose linearly such that any subsample of query instances can provide an unbiased gradient estimate.
Note that this approach works for any differentiable weighting function $\weight{r}$, does not require any alternating approximations as in Majorization-Minimization, and processes each candidate document $\ypart$ including the clicked document $\ypart_i$ only once in one SGD step. 

 \begin{figure}
    \centering
    \includegraphics*[width=0.87\linewidth,trim={2.5cm 0cm 3.5cm 0.8cm},clip]{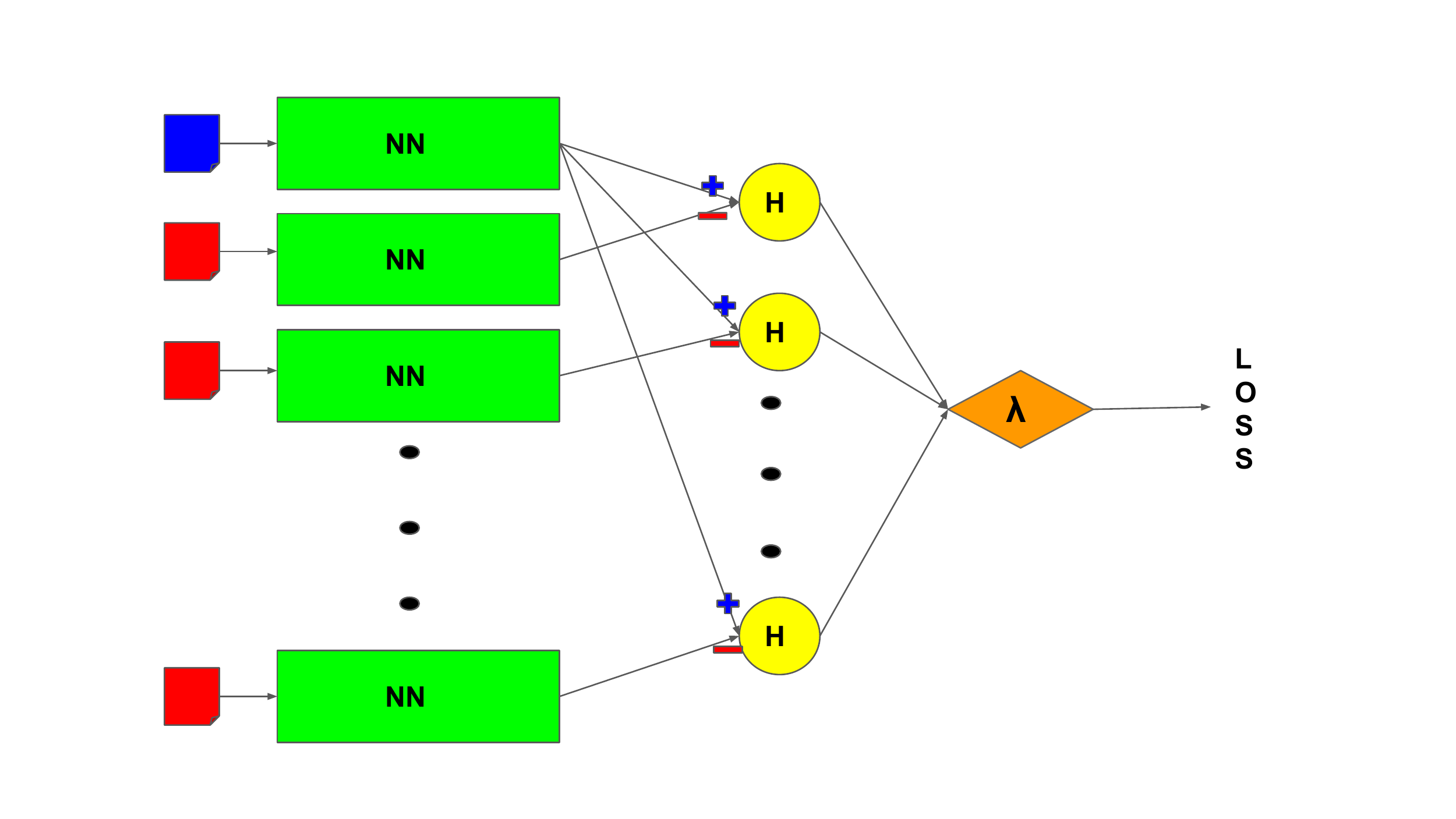}
    \vspace*{-0.3cm}
    \caption{Deep PropDCG schema for computing the loss from one query instance. The blue document is the positive (clicked) result, and the red documents are the other candidates. The neural net NN is used to compute document scores for each set of candidate features. Pairs of scores are passed through the hinge node, and then finally the weighting function is applied as shown.}
    \label{fig:nn}
\end{figure}

For SGD at the level of query instances, a forward pass of the neural network -- with the current weights fixed -- must be performed on each document $\ypart$ in candidate set $Y_i$ in order to compute the loss from training instance $(\x_i,\ypart_i)$. Since the number of documents in each candidate set varies, this is best achieved by processing each input instance (including the corresponding candidate set) as a (variable-length) sequence so that the neural net weights are effectively shared across candidate documents for the same query instance. 

This process is most easily understood via the network architecture illustrated in Figure~\ref{fig:nn}. The scoring function $NN_\w[\phi(\x_i,\ypart_i)]$ is replicated for each result in the candidate set using shared weights $\w$. In addition there is a hinge-loss node $H(u,v)=\max(1 - (u - v), 0)$ that combines the score of the clicked result with each other result in the candidate set $Y_i$. For each such pair $(\ypart_i, \ypart)$, the corresponding hinge-loss node computes its contribution $h_j$ to the upper bound on the rank. The result of the hinge-loss nodes then feeds into a single weighting node $\Lambda(\vec{h})=\weight{1+\sum_j h_j}$ that computes the overall bound on the rank and applies the weighting function. The result is the loss of that particular query instance.  

Note that we have outlined a very general method which is agnostic about the size and architecture of the neural network. As a proof-of-concept, we achieved superior empirical results over a linear scoring function even with a simple two layer neural network, as seen in Section~\ref{sec:deep}. We conjecture that DCG performance may be enhanced further with deeper, more specialized networks. Moreover, in principle, the hinge-loss nodes can be replaced with nodes that compute any other differentiable loss function that provides an upper bound on the rank without fundamental changes to the SGD algorithm. 

\section{Empirical Evaluation} 

While the derivation of SVM PropDCG and Deep PropDCG has provided a theoretical justification for both methods, it still remains to show whether this theoretical argument translates to improved empirical performance. To this effect, the following empirical evaluation addresses three key questions. 

First, we investigate whether directly optimizing DCG improves performance as compared to baseline methods, in particular, SVM PropRank as the most relevant method for unbiased LTR from implicit feedback, as well as LambdaRank, a common strong non-linear LTR method. Comparing SVM PropDCG to SVM PropRank is particularly revealing about the importance of direct DCG optimization, since both methods are linear SVMs and employ the same software machinery for the Quadratic Programs involved, thus eliminating any confounding factors. We also experimentally analyze the CCP optimization procedure to see whether SVM PropDCG is practical and efficient. Second, we explore the robustness of the generalized counterfactual LTR approach to noisy feedback, the severity of the presentation bias, and misspecification of the propensity model. And, finally, we compare the DCG performance of Deep PropDCG with a simple two layer neural network against the linear SVM PropDCG to understand to what extent non-linear models can be trained effectively using the generalized counterfactual LTR approach.  

\begin{table}[]
\begin{tabular}{lccclll}
\toprule
Dataset & \# Avg. train clicks & \# Train queries & \# Features  \\
\midrule
Yahoo  & 173,986 & 20,274  & 699\\ 
LETOR4.0 & 25,870 & 1,484  & 46\\ 
\bottomrule
\end{tabular}
\caption{Properties of the two benchmark datasets.}
\vspace{-1.5em}
  \label{tbl:datasummary}%
\end{table}

\begin{table}[]
\begin{tabular}{lll}
\toprule
Model        & Avg. DCG (Yahoo) & Avg. DCG (LETOR4.0) \\
\midrule
SVM Rank     & 0.6223 $\pm$ 8e-4 & 0.6841 $\pm$ 2e-3\\
LambdaRank   & 0.6435 $\pm$ 4e-4 & 0.6915 $\pm$ 4e-3\\
SVM PropRank & 0.6410$\pm$ 1e-3 & 0.7004 $\pm$ 1e-2 \\
SVM PropDCG  & 0.6468$\pm$ 2e-3 & 0.7043 $\pm$ 1e-2 \\
Deep PropDCG & \textbf{0.6517 $\pm$ 4e-4} & \textbf{0.7244 $\pm$ 4e-3}   \\ 
\bottomrule
\end{tabular}
\caption{Performance comparison of different methods on two benchmark datasets ($\eta = 1$, $\epsilon_-=0.1$, $\epsilon_+=1$).}
\vspace{-1.3em}
  \label{tbl:letor}%
\end{table}

\subsection{Setup} \label{setup}
We conducted experiments on synthetic click data derived from two major LTR datasets, the Yahoo Learning to Rank Challenge corpus and LETOR4.0 \cite{Qin/2013}. LETOR4.0 contains two separate corpora: MQ2007 and MQ2008. Since MQ2008 is significantly smaller than Yahoo Learning to Rank Challenge, with only 784 queries, we follow the data augmentation approach proposed in \cite{Pang/etal/17}, combining the MQ2007 and MQ2008 train sets for training and using the MQ2008  validation and test sets for validation and testing respectively. 

Our experiment setup matches \cite{Joachims/etal/17a} for the sake of consistency and reproducibility. Briefly, the training and validation click data were generated from the respective full-information datasets (with relevances binarized) by simulating the position-based click model. Following \cite{Joachims/etal/17a}, we use propensities that decay with presented rank of the result as $p_r = \big ( \frac{1}{r} \big )^\eta$. The rankings that generate the clicks are given by a ``production ranker'' which was a conventional Ranking SVM trained on 1 percent of the full-information training data. The parameter $\eta$ controls the severity of bias, with higher values causing greater position bias. 

We also introduced noise into the clicks by allowing some irrelevant documents to be clicked. Specifically, an irrelevant document ranked at position $r$ by the production ranker is clicked with probability $p_r$ times $\epsilon_-$. When not mentioned otherwise, we used the parameters $\eta = 1$, $\epsilon_-=0.1$ and $\epsilon_+=1$, which is consistent with the setup used in \cite{Joachims/etal/17a}. Other bias profiles are also explored in the following. 

Both the SVM PropRank and SVM PropDCG models were trained and cross-validated to pick the regularization constant C. For cross-validation, we use the partial feedback data in the validation set and select based on the IPS estimate of the DCG \cite{Swaminathan/Joachims/15d}. 
The performance of the models is reported on the binarized fully labeled test set which is never used for training or validation.


\subsection{How do SVM PropDCG and Deep PropDCG compare against baselines?}

We begin the empirical evaluation by comparing our counterfactual LTR methods again standard methods that follow a conventional ERM approach, namely LambdaRank and SVM-Rank. We generate synthetic click data using the procedure describe above, iterating over the training set 10 times for the Yahoo dataset and 100 times for MQ2008. This process was repeated over 6 independent runs, and we report the average performance along with the standard deviation over these runs. The regularization constant C for all SVM methods was picked based on the average DCG performance across the validation click data sampled over the 6 runs. Table~\ref{tbl:datasummary} shows the average number of clicks along with other information about the training sets.

\begin{figure}
    \centering
    \includegraphics*[width=0.87\linewidth,trim={0.5cm 0cm 0.4cm 0.2cm},clip]{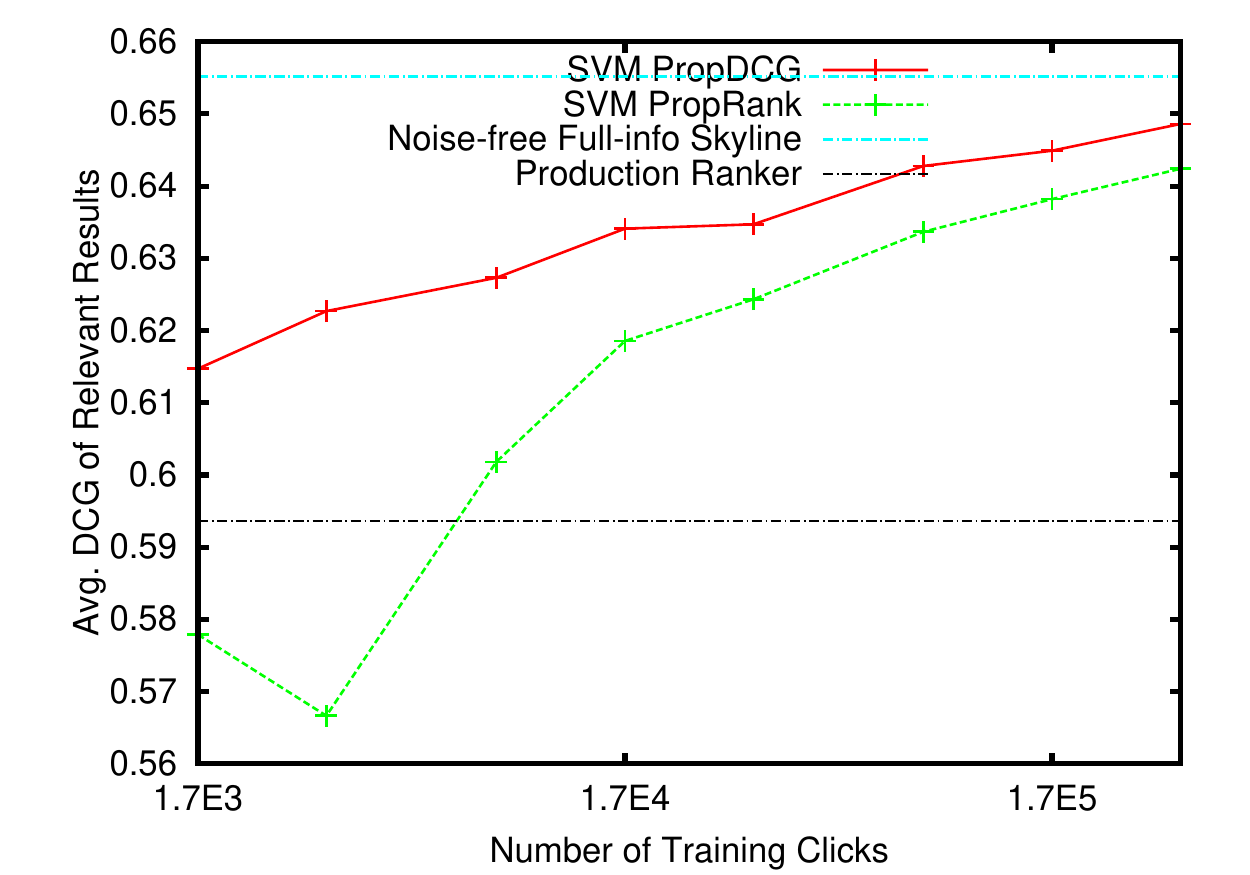}
    \vspace*{-0.3cm}
    \caption{Test set Avg DCG performance for SVM PropDCG and SVM PropRank ($\eta=1$, $\epsilon_-=0.1$)}
    \label{fig:exp_n_dcg}
\end{figure}

\begin{figure}[t]
    \centering
    \includegraphics*[width=0.87\linewidth,trim={0.5cm 0cm 0.4cm 0.4cm},clip]{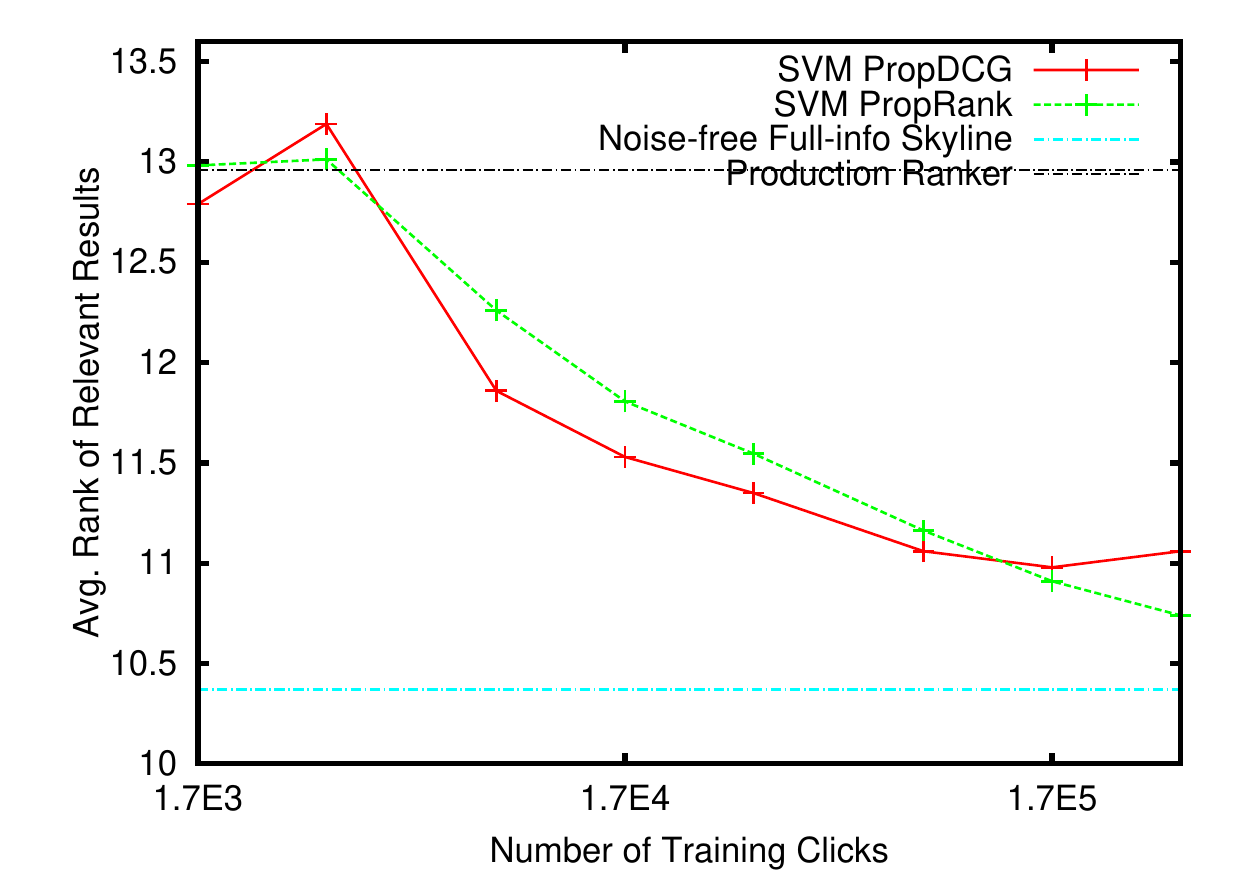}
    \vspace*{-0.3cm}
    \caption{Test set Avg Rank performance for SVM PropDCG and SVM PropRank ($\eta=1$, $\epsilon_-=0.1$)}
    \label{fig:exp_n_rank}
\end{figure}

As a representative for non-linear LTR methods that use a conventional ERM approach, we also conducted experiments with LambdaRank as one of the most popular tree-based rankers. We use the LightGBM implementation \cite{ke/etal/17}. During training, LambdaRank optimizes Normalized Discounted Cumulative Gain (NDCG). Since LambdaRank is a full-information method, we used clicks as relevance labels, i.e. all clicked documents as relevant and all non-clicked documents as irrelevant. The hyperparameters for LambdaRank, namely learning rate and the number of leaves were tuned based on the average DCG of clicked documents in the validation sets. More specifically, we performed a grid search to finetune learning rate from 0.001 to 0.1 and the number of leaves from 2 to 256. After tuning, we selected the learning rate to be 0.1, and the number of leaves to be 64 for the Yahoo dataset and 4 for MQ2008. We also made sure each split does not use more than 50\% of the input features. 




As shown in in Table~\ref{tbl:letor}, the counterfactual ERM approach via IPS weighting and directly optimizing for the target metric DCG yield superior results for SVM PropDCG and Deep PropDCG. The best results on both benchmarks are achieved by Deep PropDCG, which learns a two-layer neural network ranker. We conjecture that more sophisticated network architectures can further improve performance. 
 
\begin{figure}[t]
    \centering
    \includegraphics*[width=0.87\linewidth,trim={0.5cm 0cm 0.4cm 0.4cm},clip]{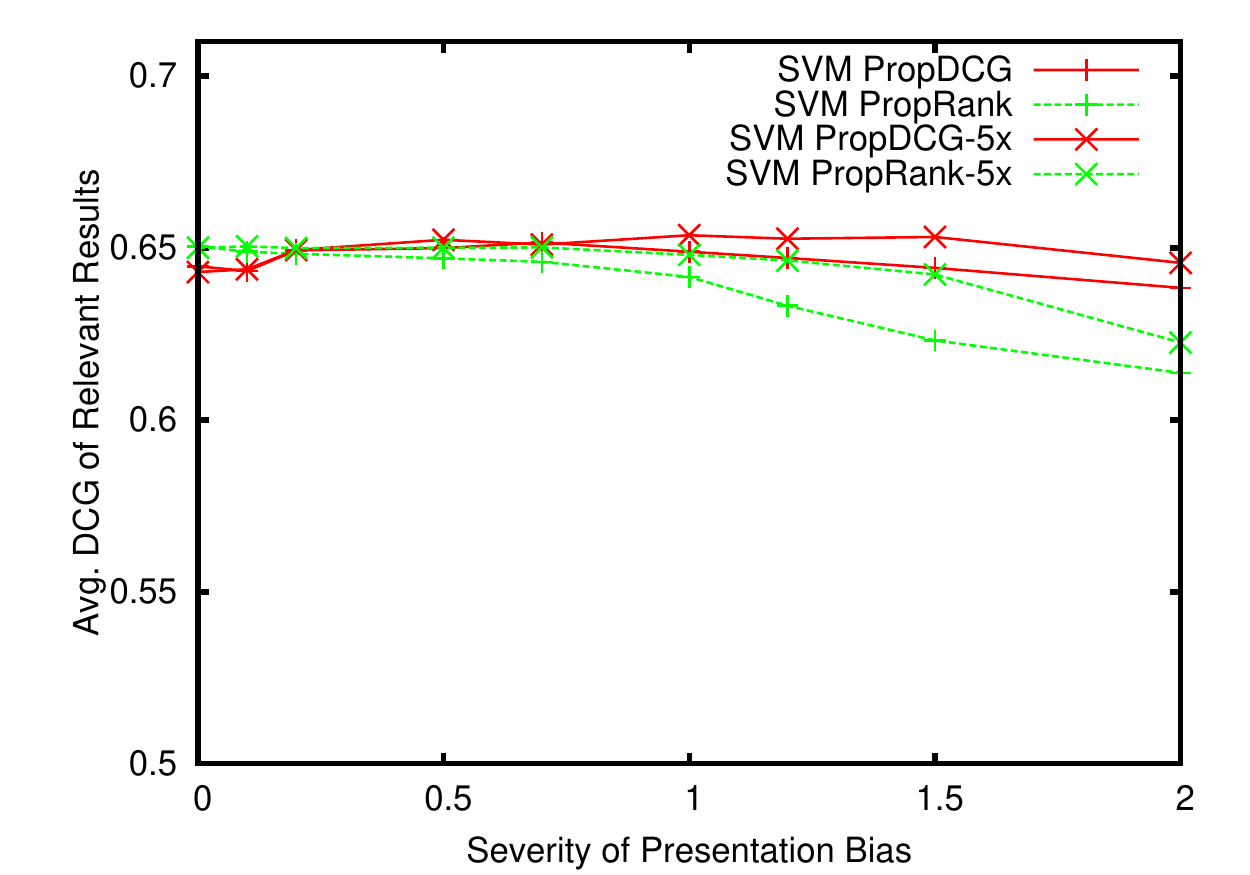}
    \vspace*{-0.3cm}
    \caption{Test set Avg DCG performance for SVM PropDCG and SVM PropRank as presentation bias becomes more severe in terms of $\eta$ ($n=45K$ and $n=225K$, $\epsilon_-=0$). }
    \label{fig:exp_prop}
\end{figure}

\subsection{How does ranking performance scale with training set size?}

Next, we explore how the test-set ranking performance changes as the learning algorithm is given more and more click data. The resulting learning curves are given in Figures~\ref{fig:exp_n_dcg} and~\ref{fig:exp_n_rank}. The click data has presentation bias with $\eta=1$ and noise with $\epsilon_-=0.1$. For small datasets, results are averaged over $3$ draws of the click data. Both curves show the performance of the Production Ranker used to generate the click data, and the SVM skyline performance trained on the full-information training set. Ideally, rankers trained on click data should outperform the production ranker and approach the skyline performance.

Figure~\ref{fig:exp_n_dcg} shows that the DCG performance of both SVM PropDCG and SVM PropRank. As expected, both improve with increasing amounts of click data. Moreover, SVM PropDCG performs substantially better than the baseline SVM PropRank in maximizing test set DCG. 

More surprisingly, Figure~\ref{fig:exp_n_rank} shows both methods perform comparably in minimizing the average rank metric, with SVM PropDCG slightly better at smaller amounts of data and SVM PropRank better at larger amounts. We conjecture that this is due the variance-limiting effect of the DCG weights in SVM PropDCG when substituting the propensity weights $q_i$ with the new constants $q_i'$ in the SVM PropDCG CCP iterations. This serves as implicit variance control in the IPS estimator similar to clipping \cite{Joachims/etal/17a} by preventing propensity weights from getting too big. Since variance dominates estimation error at small amounts of data and bias dominates at large amounts, our conjecture is consistent with the observed trend.  

\begin{figure}[t]
    \centering
    \includegraphics*[width=0.87\linewidth,trim={0.5cm 0cm 0.4cm 0.4cm},clip]{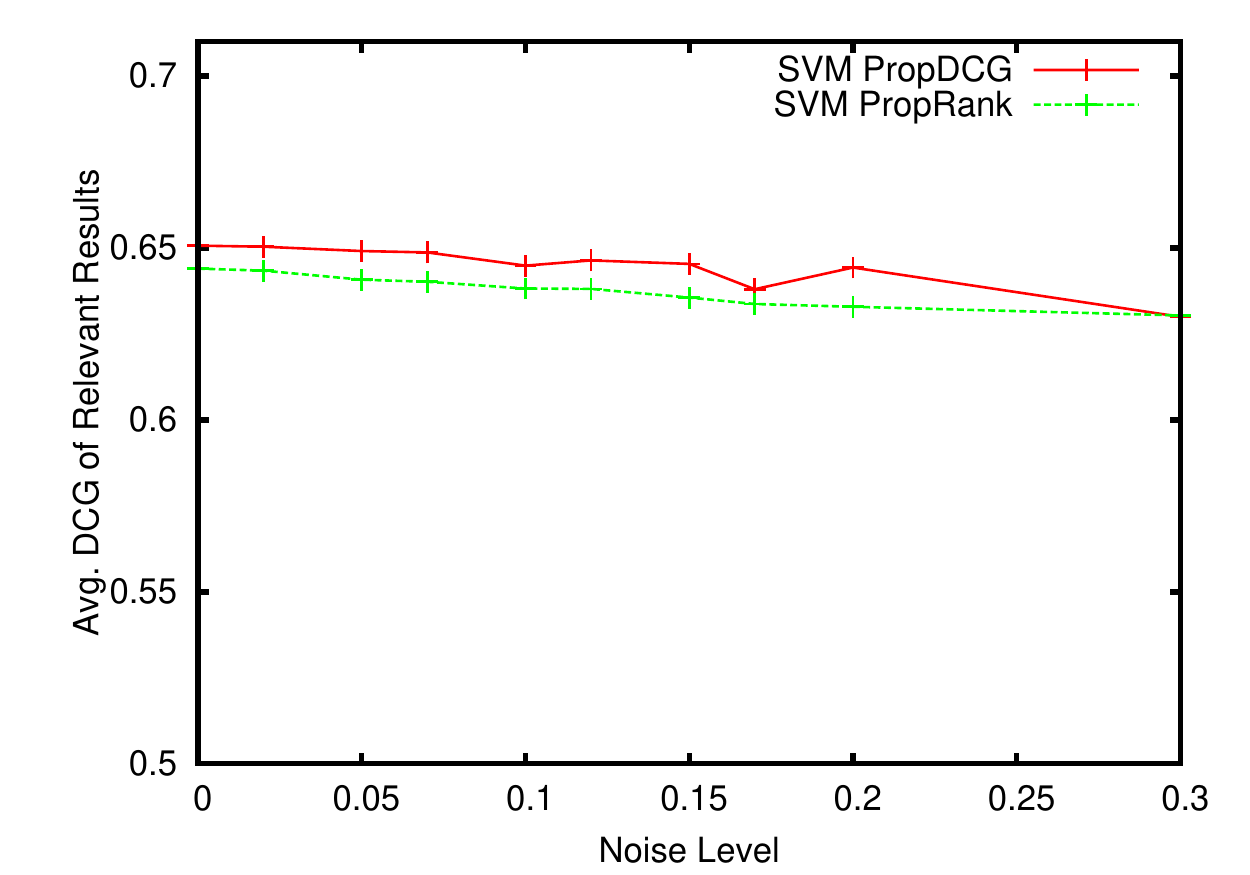}
    \vspace*{-0.3cm}
    \caption{Test set Avg DCG performance for SVM PropDCG and SVM PropRank as the noise level increases in terms of $\epsilon$ ($n=170K, \eta=1$). }
    \label{fig:exp_noise}
\end{figure}
\begin{figure}[t]
    \centering
    \includegraphics*[width=0.87\linewidth,trim={0.5cm 0cm 0.4cm 0.4cm},clip]{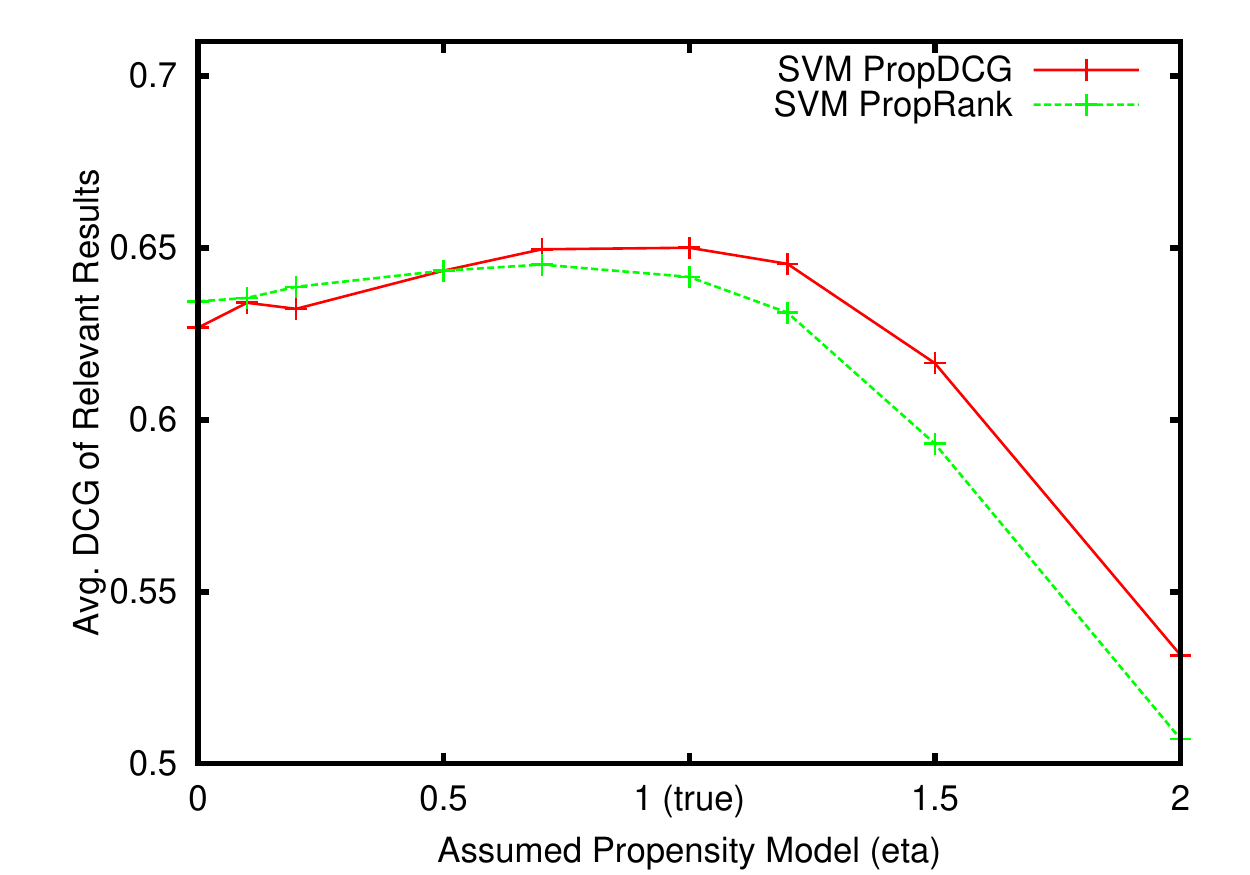}
    \vspace*{-0.3cm}
    \caption{Test set Avg DCG performance for SVM PropDCG and SVM PropRank as propensities are misspecified (true $eta=1, n=170K, \epsilon_-=0.1$). }
    \label{fig:exp_mismatch}
\end{figure}

\subsection{How much presentation bias can be tolerated?}

We now vary the severity of the presentation bias via $\eta$ -- higher values leading to click propensities more skewed to the top positions -- to understand its impact on the learning algorithm. Figure~\ref{fig:exp_prop} shows the impact on DCG performance for both methods. We report performance for two training set sizes that differ by a factor of 5 (noise $\epsilon_- = 0$). We see that SVM PropDCG is at least as robust to the severity of bias as SVM PropRank. In fact, SVM PropRank's performance degrades more at high bias than that of SVM PropDCG, further supporting the conjecture that the DCG weighting in SVM PropDCG provides improved variance control which is especially beneficial when propensity weights are large. Furthermore, as also noted for SVM PropRank in \cite{Joachims/etal/17a}, increasing the amount of training data by a factor of $5$ improves performance of both methods due to variance reduction, which is an advantage that unbiased learning methods have over those that optimize a biased objective. 
\subsection{How robust is SVM PropDCG to noise?}

Figure~\ref{fig:exp_noise} shows the impact of noise on DCG performance, as noise levels in terms of $\epsilon_-$ increase from $0$ to $0.3$. The latter results in click data where $59.8\%$ of all clicks are on irrelevant documents.
\pagebreak[4] 
As expected, performance degrades for both methods as noise increases. However, there is no evidence that SVM PropDCG is less robust to noise than the baseline SVM PropRank. 

\subsection{How robust is SVM PropDCG to misspecified propensities?}
So far all experiments have had access to the true propensities that generated the synthetic click data. However, in real-world settings propensities need to be estimated and are necessarily subject to modeling assumptions. So, we evaluate the robustness of the learning algorithm to propensity misspecification.

Figure~\ref{fig:exp_mismatch} shows the performance of SVM PropDCG and SVM PropRank when the training data is generated with $\eta=1$, but the propensities used in learning are misspecified according to the $\eta$ on the x-axis. The results show that SVM PropDCG is at least as robust to misspecified propensities as SVM PropRank. Both methods degrade considerably in the high bias regime when small propensities are underestimated -- this is often tackled by clipping \cite{Joachims/etal/17a}. It is worth noting that SVM PropDCG performs  better than SVM PropRank when misspecification leads to propensities that are underestimated, further strengthening the implicit variance control conjecture for SVM PropDCG discussed above. 

\begin{figure*}[t]
    \centering
    \includegraphics[width=0.49\textwidth]{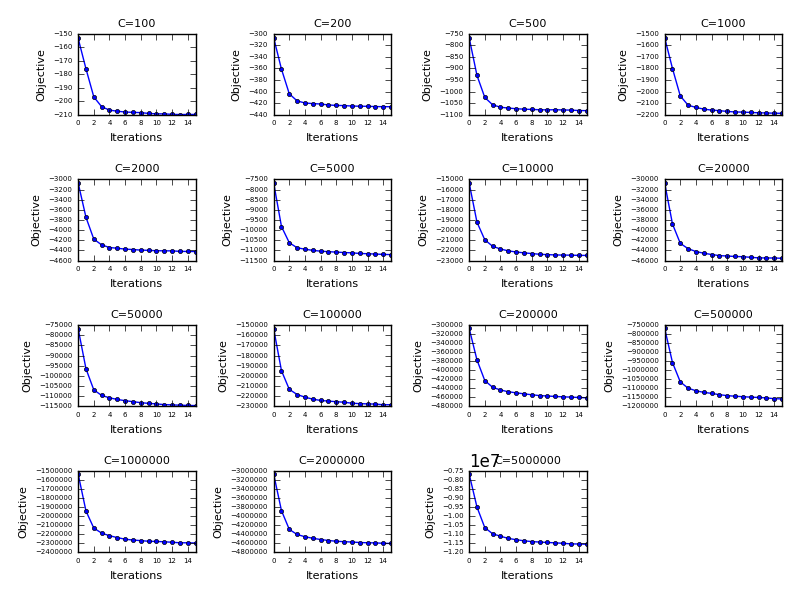} \hfill 
    \includegraphics[width=0.49\textwidth]{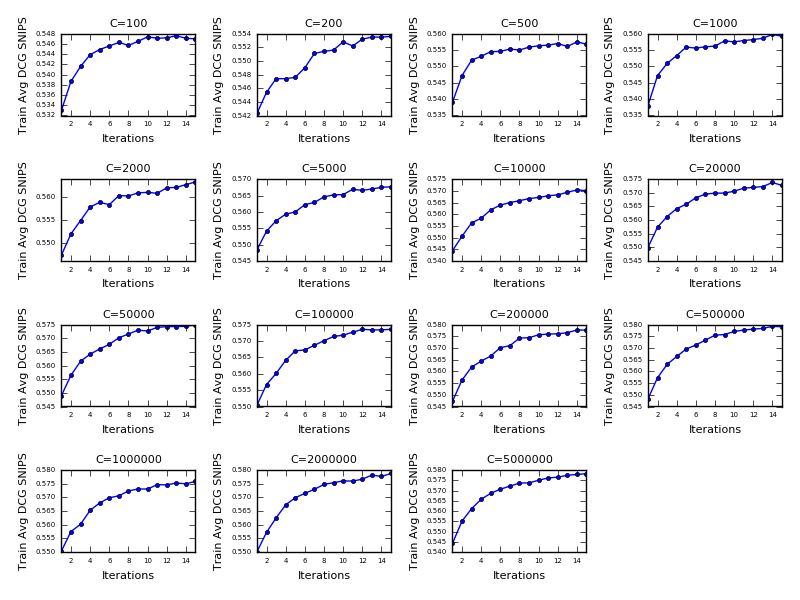} \hfill
    \caption{Optimization progress with respect to the number of CCP iterations. The objective value is shown in the left plots, and the training set DCG estimate on the right plots. Each plot corresponds to a particular value of regularization constant $C$ ($n=17K$, $\eta=1$, $\epsilon_-=0.1$).}
    \label{fig:ccp}
\end{figure*}

\subsection{How well does the CCP converge?}

Next, we consider the computational efficiency of employing the CCP optimization procedure for training SVM PropDCG. Recall that the SVM PropDCG objective is an upper bound on the regularized (negative) DCG IPS estimate. It is optimized via CCP which repeatedly solves convex subproblems using the SVM PropRank solver until the objective value converges. 

In Figure~\ref{fig:ccp}, optimization progress vs number of iterations as indicated by the change in objective value as well as the training DCG SNIPS estimate \cite{Swaminathan/Joachims/15d} is shown for $17K$ training clicks and the full range of regularization parameter $C$ used in validation. The figure shows that the objective value usually converges in 3-5 iterations, a phenomenon observed in our experiments for other amounts of training data as well. In fact, the convergence tends to take slightly fewer iterations for larger amounts of data. The figure also shows that progress in objective is well-tracked with progress in the training DCG estimate, which suggests that the objective is a suitable upper bound for DCG optimization.  

It is worth noting that restarting the optimizer across multiple CCP iterations can be substantially less time consuming than the initial solution that SVM PropRank computes. Since only the coefficients of the Quadratic Program change, the data does not need to be reloaded and the optimizer can be warm-started for quicker convergence in subsequent CCP iterations.


\subsection{When does the non-linear model improve over the linear model?}
\label{sec:deep}

\begin{figure}
    \centering
    \includegraphics*[width=0.87\linewidth,trim={0.5cm 0cm 0.4cm 0.2cm},clip]{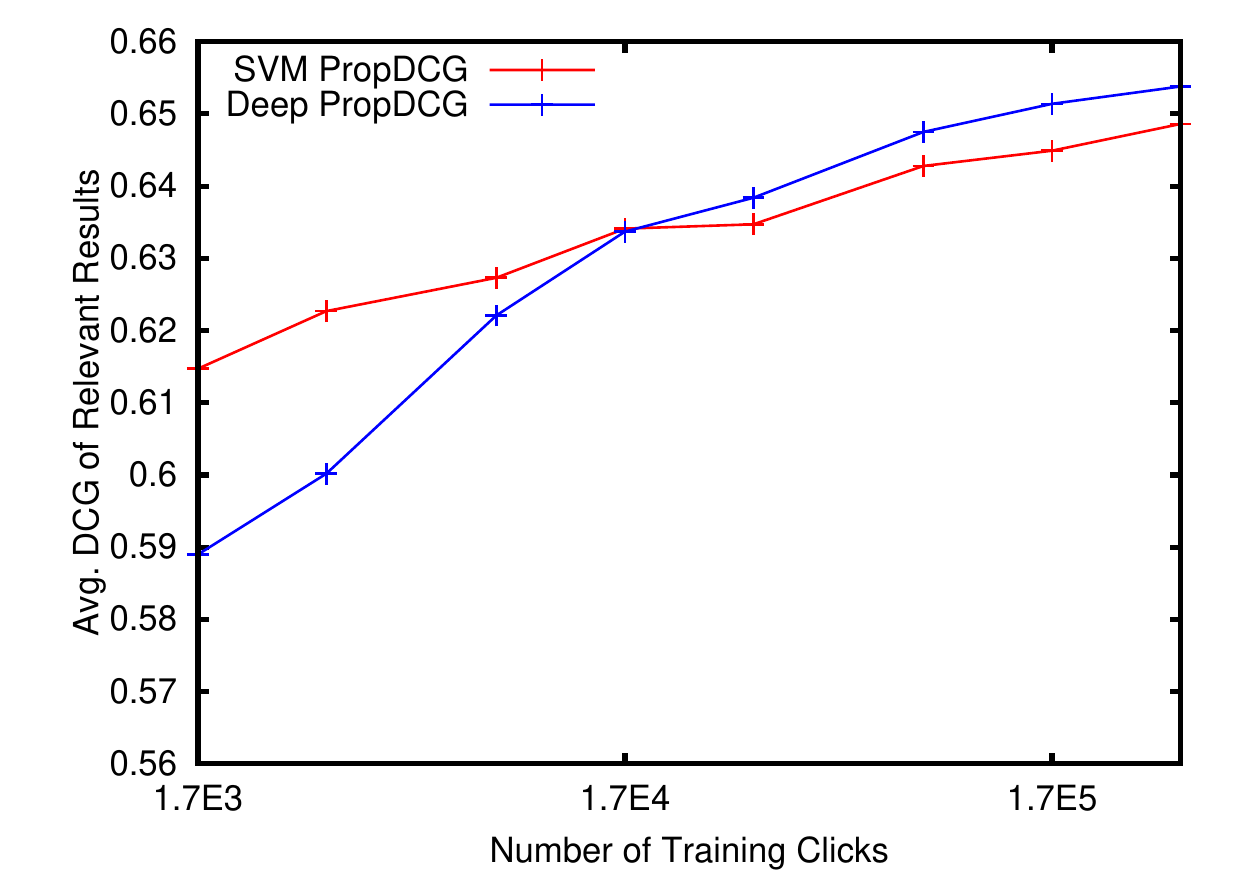}
    \vspace*{-0.3cm}
    \caption{Test set Avg DCG performance for SVM PropDCG and Deep PropDCG ($\eta=1$, $\epsilon_-=0.1$) }
    \label{fig:exp_n_dcgdeep}
\end{figure}

We have seen that SVM PropDCG optimizes DCG better than SVM PropRank, and that it is a robust method across a wide range of biases and noise levels. Now we explore if performance can be improved further by introducing non-linearity via neural networks. Since the point of this paper is not a specific deep architecture but a novel training objective, we used a simple two-layer neural network with 200 hidden units and sigmoid activation. We expect that specialized deep architectures will further improve performance.
 
Figure~\ref{fig:exp_n_dcgdeep} shows that Deep PropDCG achieves improved DCG compared to the linear SVM PropDCG given enough training data. For small amounts of training data, the linear model performs better, which is to be expected given the greater robustness to overfitting of linear models. 
 
We also expect improved performance from tuning the hyperparameters of Deep PropDCG. In fact, we only used default parameters for Deep PropDCG, while we optimized the hyperparameters of SVM PropDCG on the validation set. In particular, Adam was used for stochastic gradient descent with weight decay regularizer at $10^{-6}$, minibatch size of $1000$ documents and $750$ epochs. The learning rate began at $10^{-6}$ for the first $300$ epochs, dropping by one order of magnitude in the next $200$ epochs and another order of magnitude in the remaining epochs. We did not try any other hyperparameter settings and these settings were held fixed across varying amounts of training data. 

\section{Conclusion}

In this paper, we proposed a counterfactual learning-to-rank framework that is broad enough to cover a broad class of additive IR metrics as well as non-linear deep network models. Based on the generalized framework, we developed the SVM PropDCG and Deep PropDCG methods that optimize DCG via the Convex-Concave Procedure (CCP) and stochastic gradient descent respectively. We found empirically that SVM PropDCG performs better than SVM PropRank in terms of DCG, that it is robust to a substantial amount of presentation bias, noise and propensity misspecification, and that it can be optimized efficiently. DCG was improved further by using a neural network in Deep PropDCG. 

There are many directions for future work. First, it is open for which other ranking metrics it is possible to develop efficient and effective methods using the generalized counterfactual framework. Second, the general counterfactual learning approach may also provide unbiased learning objectives for other settings beyond ranking, like full-page optimization and browsing-based retrieval tasks. Finally, it is an open question whether non-differentiable (e.g. tree-based) ranking models can be trained in the counterfactual framework as well.

\section{Acknowledgments}

This research was supported in part by NSF Awards IIS-1615706 and IIS-1513692, an Amazon Research Award, and the Criteo Faculty Research Award program. All content represents the opinion of the authors, which is not necessarily shared or endorsed by their respective employers and/or sponsors.



\bibliographystyle{ACM-Reference-Format}
\balance
\bibliography{ref} 

\end{document}